# ULTRAFAST TEMPERATURE PROFILE CALCULATION IN IC CHIPS

*Travis Kemper, Yan Zhang, Zhixi Bian and Ali Shakouri*

Baskin School of Engineering, University of California Santa Cruz
1156 High Street, Santa Cruz, CA 95064, U.S.A.

**ABSTRACT**
One of the crucial steps in the design of an integrated circuit is the minimization of heating and temperature non-uniformity. Current temperature calculation methods, such as finite element analysis and resistor networks have considerable computation times, making them incompatible for use in routing and placement optimization algorithms. In an effort to reduce the computation time, we have developed a new method, deemed power blurring, for calculating temperature distributions using a matrix convolution technique in analogy with image blurring. For steady state analysis, power blurring was able to predict hot spot temperatures within 1°C with computation times 3 orders of magnitude faster than FEA. For transient analysis the computation times where enhanced by a factor of 1000 for a single pulse and around 100 for multiple frequency application, while predicting hot spot temperature within about 1°C. The main strength of the power blurring technique is that it exploits the dominant heat spreading in the silicon substrate and it uses superposition principle. With one or two finite element simulations, the temperature point spread function for a sophisticated package can be calculated. Additional simulations could be used to improve the accuracy of the point spread function in different locations on the chip. In this calculation, we considered the dominant heat transfer path through the back of the IC chip and the heat sink. Heat transfer from the top of the chip through metallization layers and the board is usually a small fraction of the total heat dissipation and it is neglected in this analysis.

**INTRODUCTION**

As power densities in integrated circuits (IC's) increases over time, controlling temperature is becoming an increasing integral aspect of the IC design [1]. With the increasing clock frequencies and shrinking die size, the temperature increase in IC chips can cause significant problems. The sub-threshold leakage power grows exponentially with temperature. Interconnect delay increases 5-8% per 10°C temperature increase. Cross-talk noise level increases on the order of 25% when there's 10°C temperature difference between an aggressor and a victim. Furthermore, high temperature operation reduces the lifetime of an IC. Both electromigration and oxide breakdown are exponential functions of temperature. Compounding the problem is the fact that the cost of an IC's thermal package also increases by $1 per Watt dissipated when powers are over 35-40W [2]. Therefore, efficient thermal design is becoming increasingly essential given the current trend in IC power consumption.

In order to evaluate the thermal design of a proposed IC configuration, it is imperative to theoretically predict the temperature distribution of the active elements during operation to determine the maximum temperatures the chip will endure, in order to exclude designs that exceed the critical junction temperature. This analysis is further complicated considering the spatial and temporal non-uniform heating found in contemporary microprocessors, resulting in certain regions exceeding the mean operating temperature, which are known as "hot spots" [2]. Therefore, it is necessary to model a full temperature map of the IC's surface under steady state and transient conditions for thermal analysis to identify the extreme temperatures produced by hot spots. Current methods, such as finite element analysis (FEA), use the differential heat conduction equation,

$$\rho(\vec{x})c(\vec{x})\frac{\partial T(\vec{x},t)}{\partial t} = H(\vec{x},t) + \nabla_{\vec{x}} \cdot (\lambda(\vec{x})\nabla_{\vec{x}} T(\vec{x},t)) \quad (1)$$

where, T, H, ρ, c, and λ denote temperature, heat flux, mass density, specific heat and thermal conductivity, respectively. From a specified power map, the resulting temperature distribution can be calculated [3,4,5]. While this method is accurate, it is too time consuming to be implemented for large number of nodes representing a realistic IC and also to be used in routing and placement optimization algorithms. Therefore, an alternative method, with much faster calculation speeds, is necessary to ensure proper thermal design.

Consequently, there has been a considerable amount of effort put forth in devising a faster method [6,7,8,9].

One proposed method is to model the IC as a network of thermal resistors and capacitors. This method has been used to accurately predict steady temperature distributions in VLSI systems in order to reduce hot spot temperatures [10]. However, this method is limited by the considerable effort necessary to model each sub-region of an IC's surface as a circuit consisting of connections through the package to the heat sink, where each layer's thermal resistance must be found analytically or experimentally [6]. A further limitation of this method is the substantial calculation times associated with solving a complex network with large number of nodes.

In this paper, we present a fast and accurate numerical method to calculate the surface temperature of an IC chip. This method is based on optical filtering technique in image processing. Blurred version of an image can be produced based on a known point spread function. After a review of the optical blurring, we will apply this technique to power dissipation map on a chip.

## OPTICAL BLURRING

A black and white digital image consists of a matrix of numerical representations of tonal value. The process of spatial filtering replaces the value of each pixel in an image with a new value in order to achieve certain goals, such as sharpening or blurring. In the case of blurring an image, *f*, is convoluted with another matrix *w*, known as a mask, to produce the blurred image *g*, by the follow equation

$$g(x,y) = \sum_{s=-a}^{a} \sum_{t=-b}^{b} w(s,t) \cdot f(x+s, y+t) \quad (2)$$

where *a=(m-1)/2* and *b=(n-1)/2* for a *m x n* mask [11].

In this paper we use the blurring of tonal value in accordance to Equation 2 in order to model the heat transfer process on the surface of an IC, hence the name power blurring. Power dissipation on the surface of an IC is represented by a power matrix and convoluted with a mask representing the heat spreading function. The results gave us the temperature map on the chip. The convolution could be done in frequency domain as multiplication. There are several papers related to fast thermal simulations using *Green's function technique* [5-9]. Our approach can be described in terms of the Green's function since we use the impulse response of the system and then we do a convolution. However, in the existing literature, analytical expressions of the Green function are used. The problem is that the accurate analytical solutions exist only for very simplified geometries and it is hard to obtain analytical solutions for realistic complicated packages. The analogy with image blurring, allows us to use a parameterized point spread function. This can be calculated with full 3D FEA simulations for a point source and then applied to an arbitrary power dissipation map quite accurately. This is extremely useful when different floorplans of the IC chip are investigated, or when temperature-aware routing and placement optimization is done. Since the major part of the heat spreading happens in the silicon substrate, the shape of the point spreading function does not change very much for different packages. The main differences are in the amplitude and in the width of the point spread function that could be easily parameterized for different configurations.

## FINITE ELEMENT ANALYSIS

A flip chip package was simulated using ANSYS FEA software. [3] This model consisted of a silicon IC with a surface area of 1 x 1 cm and a Cu heat sink, with an intermittent heat spreading layer surrounded by a package lid, as illustrated in Fig. 1. The die was orthogonally meshed with congruent elements 0.025 x 0.025cm, to facilitate matrix representation. Using this model, heat fluxes were applied to the IC surface to simulate heating resulting from active transistors. Materials properties are listed in Table 1. In this calculation, we considered the dominant heat transfer path through the back of the IC chip and the heat sink. Heat transfer from the top of the chip through metallization layers and the board is usually a small fraction of the total heat dissipation and it is neglected in this analysis.

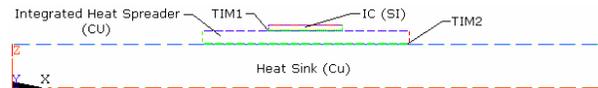

**Figure 1** FEA package model

**Table 1 Material properties**

(TIM stands for thermal interface material)

| Material | Thermal conductivity (W/m-K) | Density (Kg/m^3) | Specific heat (J/Kg-K) |
|---|---|---|---|
| Silicon | 117.5 | 2330 | 700 |
| Copper | 395 | 8933 | 397 |
| TIM1 | 5.91 | 1930 | 15 |
| TIM2 | 3.5 | 1100 | 1050 |

## POWER BLURRING

To create a mask, the temperature impulse response of the system is determined via the application of an approximate delta function power using ANSYS. In that, a *6,250 W/cm²* heat flux was applied to a single (*6.25x10⁴ cm²*) element in the center of the simulated IC, to achieve a steady-state temperature distribution. The resulting temperature distribution was normalized by the amount of power applied to produce a mask, with units of *°C/W*.

Considering that the temperature spreading function is different when the element is near the edge of the chip, a correction was applied. This position dependent scaling function can be found by using FEA or by simple analytical approximations due to dominant heat conduction in silicon substrate.

In order to test the validity of the power blurring method, we use it to analyze the power map from a published paper illustrated in Figure 2. [14]

The error of the power blurring method was defined for each element on the chip surface by the following expression:

$$Error(x, y) = \frac{PB(x, y) - FEA(x, y)}{FEA(x, y)} \quad (3)$$

where *PB(x,y)* is the temperature of the element *(x,y)* predicted by power blurring and *FEA(x,y)* is the temperature on that element as predicted by FEA, using ANSYS.

The steady state solution was obtained using FEA and power blurring. A cross section of the temperature profile along the diagonal axis is shown in Fig. 3 a-b. We can see that power blurring is able to accurately predict the temperature profile within 0.8°C. Furthermore, the FEA had a computation time of 95 seconds, while power blurring had a computation time of 0.031 seconds. Therefore, the power blurring method increased steady state computation time by a factor of more than 3000.

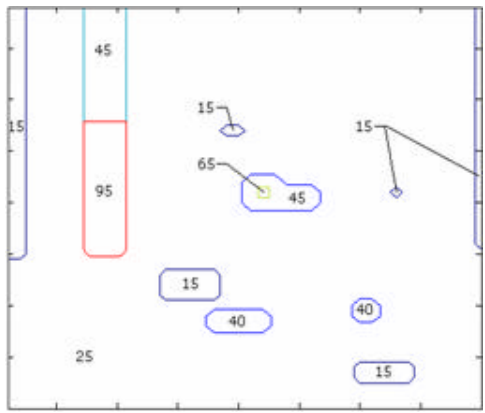

**Figure 2:** Power dissipation in W/cm² on a 1 x 1 cm IC surface.

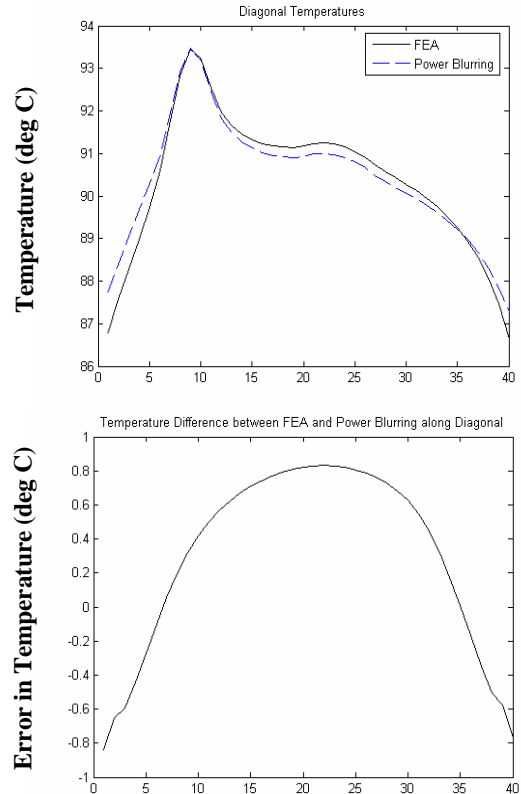

**Figure 3:** (a) Calculated temperature profile along the diagonal axis using FEA and power blurring techniques, (b) Power blurring's error in calculating the temperature.

In transient analysis, two situations were analyzed, a single pulse excitation and multiple frequency excitations. For the single pulse analysis, a square wave power dissipation with a period of 1 second was applied to the entire surface of the chip and the whole power map shown in Fig. 2 was turned on at time t=0 and turned off at time t=0.5 seconds. 25ms time step was used in the calculation of the transient temperature profile. As illustrated in Fig.4, the maximum difference between hot spot temperature calculated with FEA and with power blurring technique was 0.39°C. Furthermore, the FEA had a computation time of 4911 seconds, while power blurring had a computation time of 7 seconds. Therefore, the power blurring method increased steady state computation time by a factor of ~700.

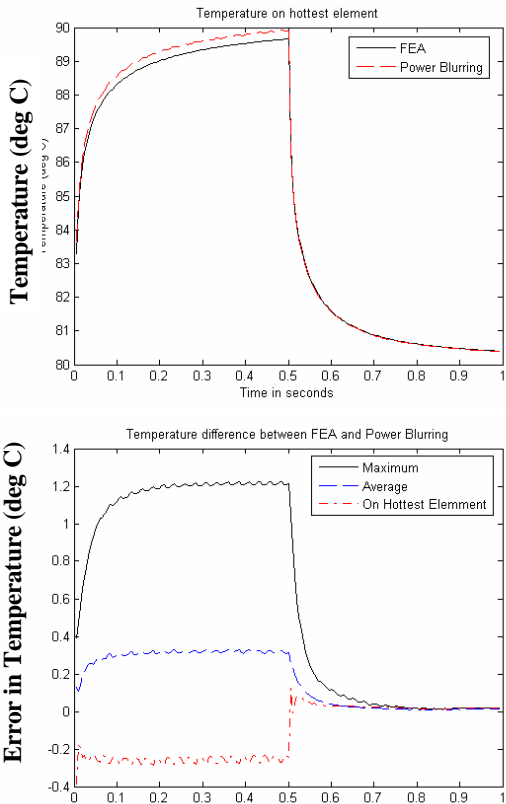
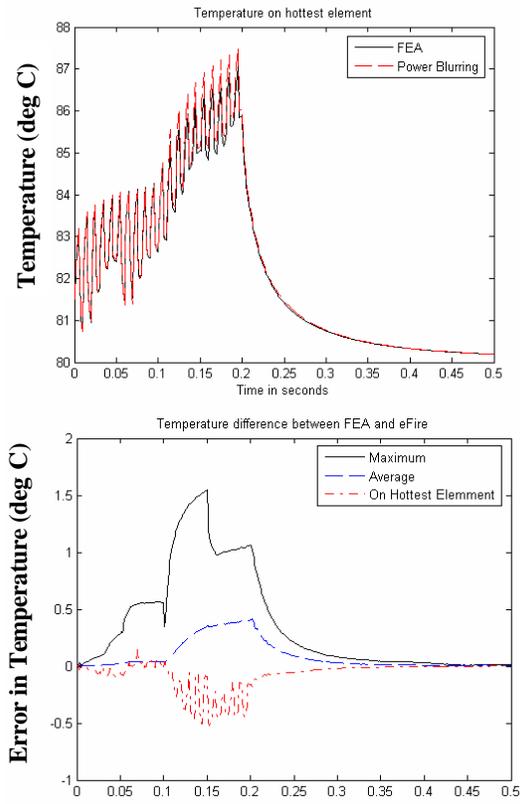

**Figure 4** (a) Calculated transient temperature profile at the hot spot using FEA and power blurring techniques (b) Power blurring's error in calculating the transient temperature response on the hottest element of the chip, on the average for the whole chip and for the element with maximum error.

**Figure 5** (a) Calculated transient temperature profile at the hot spot using FEA and power blurring techniques for the case where different areas of the chip are turned on and off with different frequencies, (b) Power blurring's error in calculating the transient temperature response on the hottest element of the chip, on the average for the whole chip and for the element with maximum error.

For the multi-frequency case, to create a dynamic power profile different regions were pulsed with different frequencies. We pulsed 90W/cm$^2$ and 65W/cm$^2$ regions with a frequency of 100Hz. The 45 W/cm$^2$ and 40 W/cm$^2$ regions were driven at 20Hz, the 15 W/cm$^2$ region at 10Hz and the 25 W/cm$^2$ region was driven at 5Hz. As illustrated in Fig. 5, the hot spot temperature could be predicted within 0.5°C. Furthermore, the FEA had a computation time 6642 seconds, while power blurring had a computation time of 87 seconds. Therefore, the power blurring method increased multi-frequency test time by a factor of 80, while predicting the temperature everywhere within less than 1.5°C.

## SUMMARY

We have shown in this paper that power blurring method is able to produce steady-state and transient temperature distribution inside the chip comparable to that of FEA with much faster calculation times. This method is similar to the Green's function technique that has been recently applied to IC thermal analysis [5-9, 15-17]. Instead of using analytical expressions of the Green's function that could be found only for simple geometries and which require infinite series, impulse function of the package is calculated using a numerical technique and then it is applied to an arbitrary heat dissipation profile. This power blurring method can also be generalized to 3D stacked chip geometries using the superposition principle and volumetric point spread function [12,13].